# ON THE GINZBURG-LANDAU ANALYSIS OF THE UPPER CRITICAL FIELD $H_{c2}$ IN $MgB_2$

İ.N.Askerzade[1,2], A. Gencer[1,+] and N.Güçlü[1]

[1] Ankara University, Faculty of Sciences, Department of Physics,
06100-Tandoğan-Ankara/Turkey

[2] Institute of Physics, Azerbaijan Academy of Sciences, Baku, 370143-Azerbaijan

## Abstract

Temperature dependence of the upper critical field $H_{c2}$ (T) for the superconducting magnesium diboride, $MgB_2$, is studied in the vicinity of $T_c$ by using a two-band Ginzburg-Landau (G-L) theory. The temperature dependence of $H_{c2}$ (T) near $T_c$ exhibits a positive curvature. In addition, the calculated temperature dependence and its higher order derivatives are also shown to be in a good agreement with the experimental data. In analogy with the multi-band character of Eliashberg microscopic theory, the positive curvature of $H_{c2}$ (T) is described reasonably by solving the two-band of G-L theory.

[+] Corresponding Author: gencer@science.ankara.edu.tr, Tel : 90-312-212 67 20/Ext. 1161.

# I. INTRODUCTION

Recent discovery of superconductivity in an intermetallic compound of magnesium diboride, $MgB_2$, has initiated an intensive research activity (see for example Refs.[1-5]) in the condensed matter community, simply because the compound was of interest to the researches in many years. In addition to its record high critical temperature at around 40 K for a binary compound, its isotope effect was also of interest. The isotope effect was immediately studied by Bud'ko et al.[6] to test the microscopic theory of superconductivity. As a superconductor, the mechanism of superconductivity in $MgB_2$ involves interesting phenomena like two-phonon scattering and multiple gaps (Ref.[5] and the references therein), giant anharmonicity and nonlinear electron-phonon (e-ph) coupling [2].

$MgB_2$ is of a type II superconductor with the G-L parameter $\kappa=26$ as given in Ref. [7]. Meanwhile, low values of critical fields $H_{c1}$ and $H_{c2}$ are reported and the estimated values for $H_{c1}$ and $H_{c2}$ at absolute zero are about a few hundreds Oe [8] and twenty thousand Oe [9-10], respectively. Large positive curvature of the upper critical magnetic field $H_{c2}$ near critical temperature $T_c$ was reported. The similar positive curvature has also been observed experimentally in nonmagnetic borocarbides of $Y(Lu)Ni_2B_2C$ [11,12]. Microscopic mechanisms of superconductivity in $MgB_2$ and discussions of the superconducting properties implicitly involve the employment of Isotropic Single Band (ISB) model [13]. However, when analysing the superconductivity using the ISB model, the multi-band character and anisotropic Fermi surface of the superconductor have been ignored. This leads to an inadequate analysis. Although the superconductivity in $MgB_2$ is phonon mediated, it is likely that an analysis beyond the single band Eliashberg theory is needed. Hence, two-band Eliashberg theory [14] can be complementary to the ISB model. Recently two-band Eliashberg model was applied [15] for determining the behaviour of $H_{c2}$ in $MgB_2$.

Although extensive theoretical studies at microscopic level have been carried out immediately after the discovery of $MgB_2$, it is necessary to provide additional information on its superconducting properties by using macroscopic theory of G-L. In this paper, we propose that the experimental data of $H_{c2}$ for $MgB_2$ and its differentiation with respect to temperature can be described in the vicinity of $T_c$ by using a two-band macroscopic theory of G-L in a similar way used for nonmagnetic borocarbides in Ref.[16].

**II- THEORY**

We write the G-L free energy functional for two-band superconductors with two coupled superconducting order parameters of the following form similarly as in [17],

$$F_{SC} = \int d^3 r \left( F_1 + F_{12} + F_2 + H^2/8\pi \right) \qquad (1)$$

with

$$F_i = \frac{\hbar^2}{4m_i} \left| \left( \nabla - \frac{2\pi i \vec{A}}{\Phi_0} \right) \Psi_i \right|^2 + \alpha_i(T) \Psi_i^2 + \beta_i \Psi_i^2 / 2 \qquad (2)$$

$$F_{12} = \varepsilon(\Psi_1^* \Psi_2 + c.c.) + \varepsilon_1 \left\{ \left( \nabla + \frac{2\pi i \vec{A}}{\Phi_0} \right) \Psi_1^* \left( \nabla - \frac{2\pi i \vec{A}}{\Phi_0} \right) \Psi_2 + c.c. \right\} \qquad (3)$$

Here $m_i$ denotes the the effective mass of the carriers belonging to the band i (i=1,2). $F_i$ is the free energy of the separate bands. $F_{12}$ is the interaction energy term between the bands. The coefficient α depends linearly on the temperature T as $\alpha_i = \gamma_i (T - T_{ci})$, while the coefficient β is independent of temperature. γ is the proportionality constant. The quantities ε and $\varepsilon_1$ describe the interband mixing of two order parameters and their gradients, respectively. H is the external magnetic field, $\Phi_0$ is the magnetic flux quantum and $\vec{A}$ is the vector potential. In Eqns. (2) and (3), it is assumed that the order parameter $|\Psi_i|^2$ are slowly varying functions in real space. By minimization of the free energy of Eqn.(1)

$$\delta F / \delta \Psi_1^* = 0 \quad , \quad \delta F / \delta \Psi_2^* = 0 \tag{4}$$

we then obtain the usual basic equation for the description of the two-band superconductivity. For simplicity, we assume a vector potential $\vec{A} = (0, Hx, 0)$ in one dimension, then linearisation of Eqn.(4) gives

$$\frac{\hbar^2}{4m_1}(\frac{d^2}{dx^2} - \frac{x^2}{l_s^2})\Psi_1 + \alpha_1(T)\Psi_1 + \varepsilon\Psi_2 + \varepsilon_1(\frac{d^2}{dx^2} - \frac{x^2}{l_s^2})\Psi_2 = 0 \tag{5}$$

$$\frac{\hbar^2}{4m_2}(\frac{d^2}{dx^2} - \frac{x^2}{l_s^2})\Psi_2 + \alpha_2(T)\Psi_2 + \varepsilon\Psi_1 + \varepsilon_1(\frac{d^2}{dx^2} - \frac{x^2}{l_s^2})\Psi_1 = 0. \tag{6}$$

Where $l_s^2 = (\hbar c / 2eH)$ is the so called magnetic length. The coefficients $\varepsilon_1$ and $\varepsilon$ are determined by the microscopic nature of the interaction between different band electrons. Their signs can be taken arbitrarily. If the interband interaction is ignored, the Eqns. (5) and (6) are decoupled into two ordinary G-L equations with two different critical temperatures. In general, regardsless of the sign of $\varepsilon$, the superconducting phase transition occurs at a well defined temperature exceeding both $T_{c1}$ and $T_{c2}$ (see Eqn.(8)). Without linearisation, the G-L equations would include a cubic term. For analytical solutions, one can make an approximation by replacing the cubic term of Eqn.(4) by an expression of $\Psi_i^3 = \langle |\Psi_i^2| \rangle \cdot \Psi_i$. To our best of knowledge, two band G-L equations have not yet been solved by taking into account the average of the order parameter.

The upper critical field $H_{c2}$ can be obtained as the lowest eigenvalue of the equation describing two-coupled harmonic oscillators in a similar way as for a single band G-L theory [18]. Hence, we look for eigenfunctions for the equations (5,6) of the form:

$$\Psi_i = \lambda_{i1} \exp(-ax^2/2) + \lambda_{i2} \exp(-bx^2/2) \tag{7}$$

Substituting these solutions into the Eqns. (5,6), we finally obtain an equation involving the upper critical field $H_{c2}$ implicitly as:

$$\left(\alpha_1(T) + \frac{\hbar^2}{4m_1 l_s^2}\right)\left(\alpha_2(T) + \frac{\hbar^2}{4m_2 l_s^2}\right) = \left(\varepsilon - \frac{\varepsilon_1}{l_s^2}\right)^2 \qquad (8)$$

The ratio of order parameters at $T_c$ can be found from the following equations:

$$y = \frac{c_1}{c_2} = \frac{\varepsilon}{\alpha_{10}} = \frac{\alpha_{20}}{\varepsilon}, \quad \alpha_{i0} = \gamma_i(T_c - T_{c,i}) \qquad (9)$$

Using Eq.(8) one obtains for the normalized upper critical field $h_{c2} = H_{c2}/\tilde{H}_{c2}(0)$

$$h_{c2} = \left(\theta - c_0 + (A\theta^2 - B\theta + c_0^2)^{1/2}\right)a_0^{-1}. \qquad (10)$$

Where $\theta = (1 - T/T_c)$, $\tilde{H}_{c2}(0) = cT_c(\gamma_1 m_1 + \gamma_2 m_2)/\hbar e$. The curvature of the upper critical field with the dimensionless parameters reads

$$h_{c2}^{''} = a_0^{-1}(Ac_0^2 - B^2/4)/(A\theta^2 - B\theta + c_0^2)^{3/2} \qquad (11)$$

$$A = \frac{(x-1)^2}{(x+1)^2} + A_1\eta^2; \quad A_1 = 64a_1a_2\frac{x^2}{(x+1)^2}; \quad x = \frac{\gamma_1 m_1}{\gamma_2 m_2}; \qquad (12)$$

$$B = \frac{2(x-1)(a_1 x - a_2)}{(x+1)^2} + (a_1 + a_2)A_1\eta^2 + 2B_1\eta, \qquad (13)$$

$$c_0 = \frac{(a_1 x + a_2)}{(x+1)} + B_1\eta; \quad a_0 = 1 - 16x\eta^2(\varepsilon/T_c)^2/\gamma_1\gamma_2 \qquad (14)$$

$$a_i = 1 - \frac{T_{ci}}{T_c}; \quad \eta = \frac{T_c m_2 \varepsilon_1 \gamma_2}{\hbar^2 \varepsilon} \qquad (15)$$

The parameters defined in Eqns. (12-15) are all dimensionless quantities. Near $T_c$, we have an asymptotical behaviour for $h_{c2}$ of the form:

$$h_{c2} = \left((1 - \frac{B}{2c_0})\theta + \frac{A}{2c_0}\theta^2\right)a_0^{-1} \qquad (16)$$

**III- RESULTS AND DISCUSSIONS**

The experimental data for the upper critical field $h_{c2}$ of $MgB_2$ can be described with high accuracy by the simple expressions

$$h_{c2} = \frac{H_{c2}}{H_{c2}^*(0)} = \frac{\theta^{1+\alpha}}{(1-(1+\alpha)\omega + \ell\omega^2 + m\omega^3)}, \text{ with } \omega = (1-\theta)\theta^{1+\alpha} \quad (17)$$

This formula of Eqn.(17) was applied for fitting the experimental data of $h_{c2}$ in nonmagnetic borocarbides by Drechsler et al. [19]. They found an unusual positive curvature near $T_c$. Müller et al [10] have only ignored the denominator of Eqn.(17) and only used the remaining simple formula, $\theta^{1+\alpha}$ for fitting their experimental data for $MgB_2$. The critical exponent $\alpha$ determines the temperature dependence more effectively. However, in such an approximation, they did not attempt to explain the negative curvature at low temperatures. Meanwhile, the saturation (negative curvature) at low temperatures is well described by the ratio of $\ell/m$, which is assumed in analogy as sensitive to the electronic structure. For $\ell > m$ the temperature region with negative curvature becomes wide. For $\ell < m$ we have an upper critical field changing almost linearly with temperature.

In figure 1, we plot the experimental data of Bud'ko et al. [9] with open circle symbol and Müller et al. [10] with filled circle symbol for $H_{c2}(T)$ in $MgB_2$ versus the reduced temperature $T/T_c$ and also show the theoretical fits to this data by using Eqns.(10) and (17). The best fit to the experimental data was obtained by using Eqn.(17) (shown by solid lines) with fitting parameters $\mu_0 H_{c2}^*(0) = 16.2$ T, $\alpha = 0.23, \ell = 3, m = -1$. In the figure, the dotted line display the two band G-L fitting by using Eqn.(10). For this fitting, G-L parameters together with other fitting parameters are as

$$A = 0.59, B = -0.11, c_0 = 0.20, x = 3, \mu_0 \tilde{H}_{c2}(0) = 15.55 \text{ T}, T_{c1} = 20 \text{ K}, T_{c2} = 10 \text{ K}.$$

At low temperatures, $H_{c2}$ is changing almost linearly with temperature. As $T_c$ is approached from below, linearity begins to disappear giving a shallow like dependence. As shown in Fig.1, there exists a reasonable agreement between theory and experiment. Note that the single band G-L theory would only give a linear temperature dependence for $H_{c2}$. The small deviations between the data and the theory may suggest some additonal effective bands to be considered in the G-L theory to look for the exact temperature dependence of $H_{c2}$. For Müller's data, we again observe the similar temperature dependence.

In figure 2, we plot the first and second order derivatives of $h_{c2}$ of Eqn.(17) and Eqn.(11) with respect to temperature versus reduced temperature $T/T_c$. Two band G-L theory gives a negative and an almost unchanging the first derivative below $T_c$. However, as $T_c$ is aproached from below, the first derivative increases until a finite value at $T_c$. On the other hand, the first derivative of the fitting formula Eqn. (17) starts from zero at absolute zero, goes through a negative region and returns to zero at $T_c$. There is a good agreement in the temperature range from 20% to 95% of $T_c$. The plots of the second derivatives are also given in the same figure. A good agreement in the same temperature range is observed for the second derivatives, too. Note that the G-L theory holds only at temperatures close to $T_c$. Hence, we concentrate on the upper side of the curves rather than the lower side.

In the case of no intergradients of order parameter ($\eta = 0$), the G-L curvature reaches its maximum at the point $\theta = B/2A \approx 0.5$. It should be noted that a similar result (in the framework of two band BCS theory) was obtained in [20], but only for a rather high value of the mass ratio parameters x=20. The inclusion of a negative intergradient interaction shifts this maximum to the region $\theta = 0$. Physically, it means that in the vicinity of $T_c$, when both order parameters are small, their interaction becomes crucial for the behaviour of the upper critical field. As shown by Eqn. (9), the critical temperature of around $T_c = 40$ K can be obtained from $T_{c1}=20$ K and $T_{c2}=10$ K approximately with the interaction parameter $\varepsilon = 3/8$.

Upper critical field $h_{c2}$ is governed by the parameter $\varepsilon_1$. Our fitting corresponds to $\eta = -0.16$. We also take into account that the average Fermi velocity over hole tubular band is $v_{F1}=4.5 \ 10^7$ cm/s, while for the remaining last Fermi surface sheets we can choose $v_{F2}=11. \ 10^7$ cm/s. In the case of nonmagnetic borocarbides, we have the following parameters : $v_{F1}=0.85 \ 10^7$ cm/s, $v_{F2}=3.8 \ 10^7$ cm/s [14]. The theoretical fit to the experimental data of nonmagnetic borocarbides by using two band G-L theory was obtained in Ref. [16] with critical temperature parameters $T_{c1}=10$ K and $T_{c2}=3$ K, approximately to yield a value of $T_c$ about 16 K.

Note that the fitting between theory and experiement shows a deviation at temperatures very close to $T_c$(see Fig.2). It is likely that we may have two main reasons for the mismatch between theory and experiment at temperatures very close to $T_c$. First, the thermal fluctuations may enhance the positive curvature near $T_c$, which is ignored in the G-L theory as mean field theory. Second, in the solutions, we assumed a linear two band G-L theory for which there exists analytical solutions. For a better fitting, the cubic term must have been included in the two band G-L theory. Then the equations to be solve would necessiate the solutions for the vortex state in two band G-L theory. Our current interest is to seek solutions for vortex state and to map out the magnetic phase diagram in the vicinity of $T_c$ for the potential superconductor of $MgB_2$.

In summary, we have shown that the positive curvature of the upper critical field can be explained to a reasonable extent by seeking analytical solutions to the two-band G-L theory within an approximation. An exact fitting between experiement and the theory would necessitate exact solutions to the theory. However, the theory itself is phonemologic and it can explain the temperature dependence to some extent only at temperatures close to $T_c$. Further work is therefore needed to elucidate the nature of superconductivity and the interesting physical properties observed in $MgB_2$ at both microscopic level and the macroscopic level.

## IV- ACKNOWLEGMENTS

One of us (I.N.A) is grateful to graduate students of A. Kılıç and E. Aksu in the Supercondcutivity Research Group for their hospitality and discussions. This work has been financially supported in part by Ankara University Research Fund under a contract number 2000-07-05-001.

**Figure Captions**

**Figure 1.** Temperature dependence of upper critical field $H_{c2}(T)$. Open circle symbol shows the experimental data of Bud'ko et al. In Ref. [9], while filled circle symbols show the experimental data of Müller et al. In Ref. [10]. The solid and the dashed-dotted lines are the best fitting curve and the two-band theoretical calculations of Ginzburg Landau theory, respectively.

**Figure 2.** The first and second derivatives of Eqn.(10) and Eqn.(17) versus the reduced temperature. Open circle symbols show the best fit of the fitting formula of Eqn.(17), while the filled circle symbols show the derivatives by the G-L Eqn.(10).

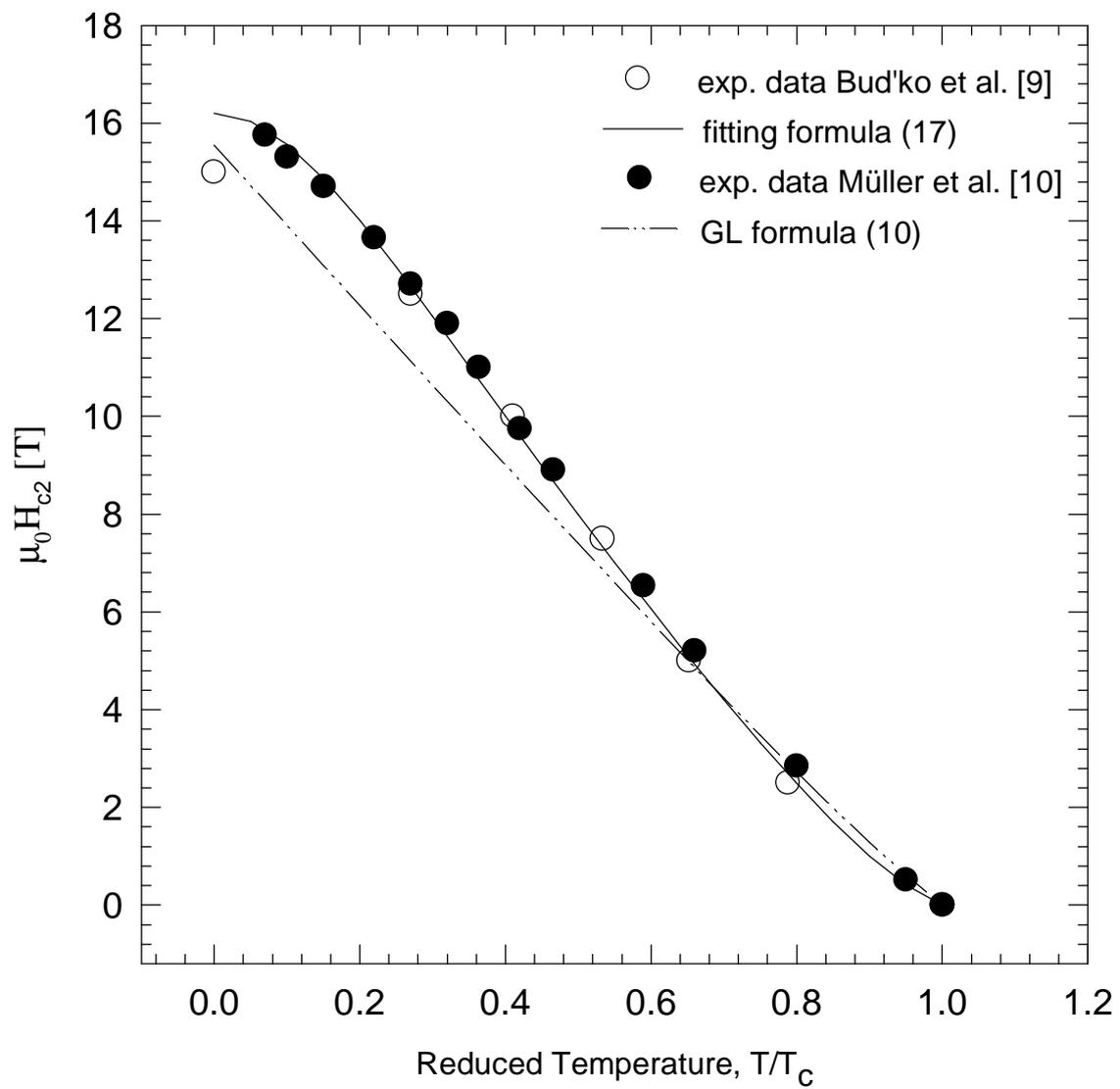

**Figure 1**

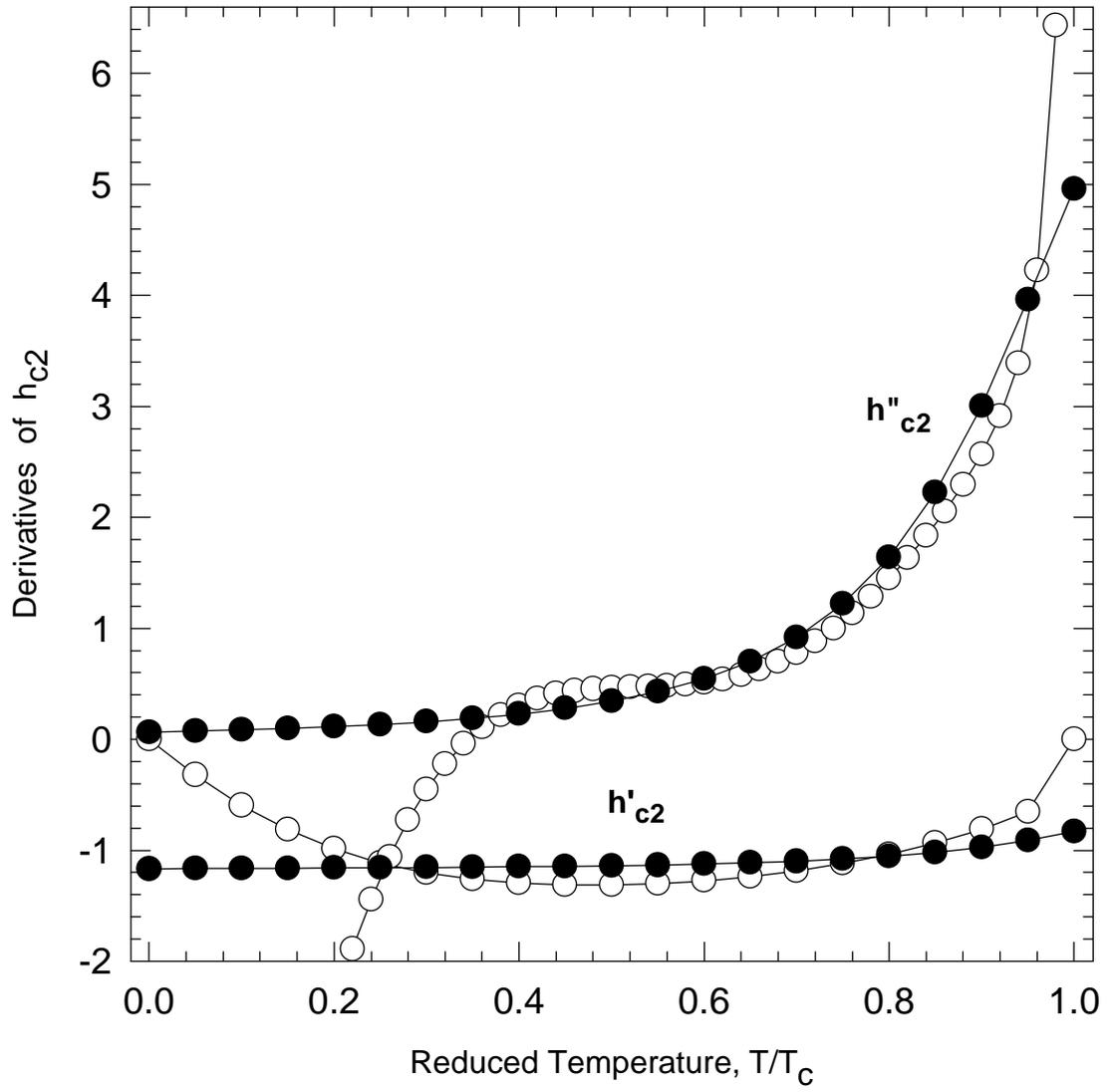

**Figure 2**